\documentclass[11pt,twoside,a4paper]{article}

\usepackage{multicol}
\usepackage{amsmath}
\usepackage{amsfonts}
\usepackage{cite}

\newcommand{\Ent}[1]{[\mkern - 2.5 mu [#1] \mkern - 2.5 mu ]}

\begin{document}

\thispagestyle{empty}

\title{Addition Theorems as Three-Dimensional Taylor Expansions. \\ 
II. $B$ Functions and Other Exponentially Decaying Functions}
\author{Ernst Joachim Weniger \\
Institut f\"{u}r Physikalische und Theoretische Chemie \\
Universit\"{a}t Regensburg \\
D-93040 Regensburg \\
Germany \\
joachim.weniger@chemie.uni-regensburg.de}
\date{Submitted to the Per-Olof L\"owdin Honorary Volume \\
International Journal of Quantum Chemistry 2002 \\
Date of Submission \\
31 August 2001}

\maketitle

\typeout{==> Abstract}
\begin{abstract} 
\noindent
Addition theorems can be constructed by doing three-dimensional Taylor
expansions according to $f (\mathbf{r} + \mathbf{r}') = 
\exp (\mathbf{r}' \cdot \mathbf{\nabla}) f (\mathbf{r})$. 
Since, however, one is normally interested in addition theorems of
irreducible spherical tensors, the application of the translation
operator in its Cartesian form $\exp (x' \partial /\partial x)
\exp (y' \partial /\partial y) \exp (z' \partial /\partial z)$ would 
lead to enormous technical problems. A better alternative consists in
using a series expansion for the translation operator $\exp (\mathbf{r}'
\cdot \mathbf{\nabla})$ involving powers of the Laplacian
$\mathbf{\nabla}^2$ and spherical tensor gradient operators
$\mathcal{Y}_{\ell}^{m} (\nabla)$, which are irreducible spherical
tensors of ranks zero and $\ell$, respectively [F.D.\ Santos, Nucl.\
Phys.\ A {\bf 212}, 341 (1973)]. In this way, it is indeed possible to
derive addition theorems by doing three-dimensional Taylor expansions
[E.J.\ Weniger, Int.\ J.\ Quantum Chem.\ {\bf 76}, 280 (2000)].  The
application of the translation operator in its spherical form is
particularly simple in the case of $B$ functions and leads to an
addition theorem with a comparatively compact structure. Since other
exponentially decaying functions like Slater-type functions, bound-state
hydrogenic eigenfunctions, and other functions based on generalized
Laguerre polynomials can be expressed by simple finite sums of $B$
functions, the addition theorems for these functions can be written down
immediately.
\end{abstract}

\begin{multicols}{2}

\typeout{==> Section 1}
\setcounter{equation}{0}

\section{Introduction}
\label{Sec:Intro}

In many branches of physics and physical chemistry -- for example in
electrodynamics \cite{Jac1975}, in classical field theory
\cite{Jon1995}, or in the theory of intermolecular forces \cite{Sto1996} 
-- an essential step towards a solution of the problem under
consideration consists in a \emph{separation} of the variables. 

Principal tools, which can accomplish such a separation of variables,
are \emph{addition theorems}: These are expansions of a given function
$f (\mathbf{r} \pm \mathbf{r}')$ with $\mathbf{r}, \mathbf{r}'
\in \mathbb{R}^3$ in terms of other functions that only depend on either
$\mathbf{r}$ or $\mathbf{r}'$. 

In atomic and molecular calculations, one is usually interested in
irreducible spherical tensors which can be partitioned into a radial
part and an angular part that is given by a spherical harmonic:
\begin{equation}
F_{\ell}^{m} (\mathbf{r}) \; = \;
f_{\ell} (r) \, Y_{\ell}^{m} (\mathbf{r}/r) \, .
\label{IrrSpheTen}
\end{equation}  
Thus, the function $f (\mathbf{r} \pm \mathbf{r}')$, which is to be
expanded, as well as the functions, that only depend on either
$\mathbf{r}$ or $\mathbf{r}'$, should be irreducible spherical tensors. 

The best known example of such an addition theorem is the Laplace
expansion of the Coulomb potential in terms of spherical harmonics:
\begin{eqnarray}
\lefteqn{\frac{1}{\vert {\mathbf{r}} - {\mathbf{r}}' \vert} \; = \;
\sum_{\ell=0}^{\infty} \, \sum_{m=-\ell}^{\ell} \,
\frac{4 \pi}{2 \ell + 1}} \nonumber \\
& \times & \! \! \frac{r_{<}^{\ell}}{r_{>}^{\ell+1}} \,
\bigl[ Y_{\ell}^{m} ({\mathbf{r}_{<}}/r_{<}) \bigr]^{*} \,
Y_{\ell}^{m} ({\mathbf{r}}_{>}/r_{>}) \, , \nonumber \\
\lefteqn{\; r_{<} = \min (r,r')\, , \; r_{>} = \max (r,r') \,
.\: \;}
\label{LapExp}
\end{eqnarray}   
The Laplace expansion leads to a separation of the variables
${\mathbf{r}}$ and ${\mathbf{r}}'$. However, the right-hand side of this
expansion depends on ${\mathbf{r}}$ and ${\mathbf{r}}'$ only indirectly
via the vectors $\mathbf{r}_{<}$ and $\mathbf{r}_{>}$ which satisfy
$\vert \mathbf{r}_{<} \vert < \vert \mathbf{r}_{>} \vert$. Hence, the
Laplace expansion has a \emph{two-range form}, depending on the relative
size of $\mathbf{r}$ and $\mathbf{r}'$. This is a complication which
occurs frequently among addition theorems.

There is an extensive literature on addition theorems. Since addition
theorems can be viewed as expansions in terms of spherical harmonics
with arguments ${\mathbf{r}}/r$ and ${\mathbf{r}}'/r'$, they are often
treated in books on angular momentum theory or on related topics (see
for example pp.\ 163 - 169 of \cite{VarMosKhe1988} or Appendix H.4 of
\cite{Nor1980}).

Particularly well studied are the addition theorems of the solutions of
the homogeneous Laplace equation \cite{Hob1965,CarRus1950,Ros1958,%
Sea1961,Fon1961,Chi1964,Sac1964,DahBar1965,Ste1969,SteRue1973,Sac1974,%
Jud1975,TouSto1977,WenSte1985,Pie1986,ChaDew1995}, of the homogeneous
Helmholtz equation \cite{Jud1975,WenSte1985,FriRus1954,Ste1961,Cru1962,%
DanMax1965,Noz1966,Raf1971,SteFil1975,CleSchr1993} and of the
homogeneous modified Helmholtz equation
\cite{WenSte1985,Noz1966,Raf1971,SteFil1975,ButGol1966}. This large
number of references does not only reflect the importance of these
functions, but also the relative ease with which their addition theorems
can be derived. The derivation of addition theorems for other functions,
which are not solutions of the equations listed above, is according to
experience significantly more difficult.

In the context of atomic and molecular electronic structure
calculations, addition theorems have traditionally been used for the
separation of the variables of interelectronic repulsion integrals and
of other multicenter integrals. For that purpose, one also needs
addition theorems for the so-called basis functions. Probably the oldest
attempt is the zeta function method of Barnett and Coulson
\cite{BarCou1951} which tries to obtain addition theorems for
exponentially decaying functions by applying suitable generating
differential operators to the well-known addition theorem of the Yukawa
potential $e^{-\alpha r}/r$. In the case of Slater-type functions, this
approach did not lead to a complete success: It was not possible to do
the differentiations in closed form, and the coefficients occurring in
the zeta function expansion had to be computed recursively
\cite{BarCou1951,Bar1963}.

Another general method for the derivation of addition theorems is the
alpha function method, which was introduced by Per-Olof L\"{o}wdin in
his seminal paper on the quantum theory of cohesive properties of solids
\cite{Low1956}, and which was later used and extended by numerous other 
authors \cite{Sha1968,Duf1971,Sha1976,SilMoa1977,JonWea1978,Ras1981,%
Ras1982,Suz1984,Ant85,Suz1985,JonBusWea1987,Suz1987,Suz1990,Suz1992,%
Jon1991,Jon1992,JonJai1996}. 

In 1967, Ruedenberg \cite{Rue1967} and Silverstone \cite{Sil1967}
suggested independently to derive addition theorems via Fourier
transformation. If the function $f (\mathbf{r} \pm \mathbf{r}')$ is an
irreducible spherical tensor of the type of (\ref{IrrSpheTen}), then the
angular integrations can be done in closed form, and there only remain
some radial integrals in momentum space containing two spherical Bessel
functions (see for example Eq.\ (7.5) of \cite{Wen1985}). Unfortunately,
these integrals can be very difficult. There is the additional
complication that even if explicit expressions for these integrals can
be found, the radial variables $r$ and $r'$ need not be separated in the
final result. Thus, the success of this approach depends crucially upon
one's ability of handling complicated integrals containing products of
spherical Bessel functions.

Addition theorems can be defined as three-dimensional Taylor expansions
(see for example p.\ 181 of \cite{BieLou1981}):
\begin{eqnarray}
f (\mathbf{r} + \mathbf{r}') & = & \sum_{n=0}^{\infty} \,
\frac{(\mathbf{r}' \cdot \mathbf{\nabla})^n}{n!} \, f (\mathbf{r})
\nonumber \\
& = & 
\mathrm{e}^{\mathbf{r}' \cdot \mathbf{\nabla}} \, f (\mathbf{r}) \, .
\label{ExpDifOp}
\end{eqnarray}
Thus, the translation operator $\mathrm{e}^{\mathbf{r}' \cdot
\mathbf{\nabla}}$ generates $f (\mathbf{r} + \mathbf{r}')$ by doing a
three-dimensional Taylor expansion of $f$ around $\mathbf{r}$. 

From a practical point of view, three-dimensional Taylor expansions do
not seem to be very useful for the derivation of addition theorems. In
atomic and molecular physics, one is usually interested in addition
theorems of irreducible spherical tensors of the type of
(\ref{IrrSpheTen}) which are defined in terms of spherical polar
coordinates $r$, $\theta$, and $\phi$. Accordingly, differentiations
with respect to the Cartesian components $x$, $y$, and $z$ of
$\mathbf{r}$ would lead to extremely messy expressions. Moreover, it is
highly desirable to express the angular part of the addition theorem in
terms of the spherical harmonics $Y_{\ell_1}^{m_1} (\mathbf{r}/r)$ and
$Y_{\ell_2}^{m_2} (\mathbf{r}'/r')$. Thus, any attempt to derive an
addition theorem for an irreducible spherical tensor via a
straightforward application of the translation operator in its Cartesian
form $\mathrm{e}^{x' \partial /\partial x}
\mathrm{e}^{y' \partial /\partial y} \mathrm{e}^{z' \partial /\partial
z}$ would lead to enormous technical problems.

Nevertheless, it is not only possible to obtain addition theorems via
three-dimensional Taylor expansions, but this approach may even be the
most natural and the most widely applicable technique
\cite{Wen2000}. This is accomplished by expanding the translation
operator $\mathrm{e}^{\mathbf{r}' \cdot \mathbf{\nabla}}$ in terms of
differential operators that are irreducible spherical tensors. Then, the
angular part of the addition theorem is obtained via angular momentum
coupling, and only radial differentiations have to be done
explicitly. Obviously, this leads to a significant simplification
\cite{Wen2000}.

In \cite{Wen2000}, it was shown how the Laplace expansion (\ref{LapExp})
of the Coulomb potential and the addition theorems of the regular and
irregular solid harmonics and of the Yukawa potential $e^{-\alpha r}/r$
can be derived by applying the translation operator in its spherical
form. Of course, these addition theorems are all comparatively simple,
and they do not necessarily provide convincing evidence that the Taylor
expansion method is indeed a useful tool for the derivation of more
complicated addition theorems. However, on p.\ 293 of \cite{Wen2000}, is
was emphasized that this method had been applied successfully also in
the case of other functions of physical interest. Thus, it is the
intention of this article to provide some evidence that these claims are
indeed true (further articles on additions theorems are in preparation
\cite{Wen????}). Moreover, in \cite{CCHiMoh2001} the addition theorem of
the so-called spherical Laguerre Gaussian functions was derived by
applying the techniques described in \cite{Wen2000}.

Currently, the vast majority of molecular electronic structure
calculations use Gaussian basis functions although these functions are
neither able to reproduce the cusps \cite{Kat1957} of the exact
solutions of molecular Schr\"odinger equations at the nuclei nor their
exponential decay \cite{Agm1982,CycFroKirSim1987}. The only, but
nevertheless decisive advantage of Gaussian functions has been the
relative ease, with which the molecular multicenter integrals can be
computed. Nevertheless, there are still many researchers who hope that
ultimately the unphysical Gaussian functions can be replaced by the
physically better motivated exponentially decaying functions and who are
doing research in this direction. Accordingly, there is a large number
of recent articles on molecular multicenter integrals of exponentially
decaying functions
\cite{BouFar1994,BouFarRin1994,BouRin1994,FeRiLopRamTab1994,BouJon1998,%
BouWeaJon1999,FeRicLopAguEmaRam1998,FeRicFerEmaLopRam2000,%
RicLopRam2001,SteHomFeRiEmaLopRam1999,SteHomEmaLopRam2000,SafPinHog1998,%
SafHog1998,SafHog1999a,SafHog1999b,SafHog1999c,SafHog2000,Saf2000,%
Saf2001a,Saf2001b,Saf2001c,SafHog2001,BecHoh1999,MagRapCas1999,%
MagRap2000,Bar2000,NovNik200a,NovNik200b}. 

So far, Slater-type functions \cite{Sla1930} have been the most
important and most widely used exponentially decaying basis functions in
atomic, molecular, and solid state calculations. Unfortunately, the
molecular multicenter integrals of Slater-type functions are notoriously
difficult, and so is its addition theorem which can be used for the
separation of variables in multicenter integrals. Over the years, a large
variety of different techniques were used for the construction of
addition theorems for Slater-type functions
\cite{BarCou1951,Bar1963,Low1956,Sha1976,JonWea1978,Ras1981,Sil1967,%
BouJon1998,MagRapCas1999,MagRap2000,ReRiLop1986,Wea1988}, and most
likely there is still room for improvements.

The topic of this article is not the derivation of addition theorems for
Slater-type functions but for another class of exponentially decaying
functions, the so-called $B$ functions \cite{FilSte1978b}. At first
sight, this may look surprising since $B$ functions have a comparatively
complicated mathematical structure, and it is by no means obvious that
anything can be gained by considering $B$ functions instead of the
apparently much simpler Slater-type functions. However, $B$ functions
have some remarkable mathematical properties which give them a unique
position among exponentially decaying functions and make them especially
useful for molecular calculations. Due to these advantageous properties
it is not only relatively easy to derive an addition theorem with the
help of three-dimensional Taylor expansions as described in
\cite{Wen2000}, but the addition theorems for $B$ functions also have
comparatively simple structures \cite{WenSte1989}. 

Moreover, all the commonly used exponentially declining functions as for
example Slater-type functions or bound-state hydrogenic eigenfunctions
can be expressed as simple finite sums of $B$ functions (see for example
Section III of \cite{WenSte1989} and references therein). These
advantageous features have led to a considerable amount of recent
research on $B$ functions and their multicenter integrals
\cite{BouFarRin1994,FeRicFerEmaLopRam2000,%
RicLopRam2001,SteHomFeRiEmaLopRam1999,SteHomEmaLopRam2000,%
SafPinHog1998,SafHog1998,SafHog1999a,SafHog1999b,SafHog1999c,%
SafHog2000,Saf2000,Saf2001a,Saf2001b,Saf2001c,NovNik200a,NovNik200b}.

Addition theorems for $B$ functions have already been derived in
\cite{WenSte1989} by applying techniques that closely resemble the zeta 
function method of Barnett and Coulson \cite{BarCou1951}. So, the
emphasis of this article is not so much on the derivation of new results
but on the discussion of methodological questions related to the
application of the translation operator in its spherical form to
exponentially decaying functions.

In Section \ref{Sec:Defs}, the definitions and conventions used in this
article are listed. In Section \ref{Sec:TransOp}, the expansion of the
translation operator in terms of irreducible spherical tensors is
reviewed, and the properties of the differential operator
$\mathcal{Y}_{\ell}^{m} (\mathbf{\nabla})$, which is the key quantity in
this expansion, are discussed. Section \ref{Sec:BFun} covers $B$
functions and their mathematical properties, and in Section
\ref{Sec:AddTheorBFun}, an addition theorem for $B$ functions is derived
by doing a three-dimensional Taylor expansion. Finally, in Section
\ref{Sec:SumCon} there is a short discussion of the results presented in
this article.

%
\setcounter{equation}{0}

\section{Definitions}
\label{Sec:Defs}

For the commonly occurring special functions of mathematical physics we
use the notation of Magnus, Oberhettinger, and Soni \cite{MaObSo1966}
unless explicitly stated.

For the set of \emph{positive} and \emph{negative} integers, we write
$\mathbb{Z} = \{ 0, \pm 1, \pm 2, \ldots \}$, for the set of
\emph{positive} integers, we write $\mathbb{N} = \{ 1, 2, 3, \ldots \}$, 
and for the set of \emph{non-negative} integers, we write $\mathbb{N}_0 =
\{ 0, 1, 2, \ldots \}$. The real numbers are denoted by $\mathbb{R}$,
and for the set of three-dimensional vectors with real components we write
$\mathbb{R}^3$.

For the spherical harmonics $Y_{\ell}^{m} (\theta, \phi)$, we use the
phase convention of Condon and Shortley \cite{ConSho1970}, i.e., they
are defined by (p.\ 69 of \cite{BieLou1981})
\begin{eqnarray}
Y_{\ell}^{m} (\theta, \phi) & = & i^{m + \vert m \vert} \,
\left[ \frac{(2 \ell + 1)(\ell - \vert m \vert)!}
{4 \pi (\ell + \vert m \vert)!} \right]^{1/2} \nonumber \\
& & \times \, 
P_{\ell}^{\vert m \vert} (\cos \theta) \, {\mathrm{e}}^{i m \phi} \, .
\end{eqnarray}
Here, $P_{\ell}^{\vert m \vert} (\cos \theta)$ is an associated Legendre
polynomial (p.\ 155 of \cite{ConOda1980}):
\begin{eqnarray}
P_{\ell}^{m} (x) & = & (1-x^2)^{m/2} \, 
\frac{{\mathrm{d}}^{\ell+m}}{{\mathrm{d}x}^{\ell+m}} \,
\frac{(x^2-1)^{\ell}}{2^{\ell} {\ell}^{\ell}} \nonumber \\
& = & (1-x^2)^{m/2} \, 
\frac{{\mathrm{d}}^{m}}{{\mathrm{d}x}^{m}} \, P_{\ell} (x) \, .
\end{eqnarray}
Under complex conjugation, the spherical harmonics satisfy (p.\ 69 of
\cite{BieLou1981}):
\begin{equation}
\left[ Y_{\ell}^{m} (\theta, \phi) \right]^{*} \; = \;
(-1)^m \, Y_{\ell}^{-m} (\theta, \phi) \, .
\label{Ylm_CC}
\end{equation}

For the regular solid harmonics, which is a solution of the homogeneous
Laplace equation, we write:
\begin{equation}
{\mathcal{Y}}_{\ell}^{m} ({{\mathbf{r}}}) \; = \;
r^{\ell} Y_{\ell}^{m} (\theta, \phi) \, .
\label{DefYlm}
\end{equation}
The regular solid harmonic is a \emph{homogeneous} polynomial of degree
$\ell$ in the Cartesian components $x$, $y$, and $z$ of $\mathbf{r}$
(p.\ 71 of
\cite{BieLou1981}):
\begin{eqnarray}
\lefteqn{\mathcal{Y}_{\ell}^{m} (\mathbf{r}) \; = \; 
\left[ \frac{2\ell+1}{4\pi} (\ell+m)!(\ell-m)! \right]^{1/2}} 
\nonumber \\
& & \times \, \sum_{k \ge 0} 
\, \frac
{(-x-iy)^{m+k} (x-iy)^{k} z^{\ell-m-2k}}
{2^{m+2k} (m+k)! k! (\ell-m-2k)!} \, . \qquad
\label{YlmHomPol} 
\end{eqnarray}
This formula holds for the Cartesian components of an arbitrary
three-dimensional vector. Consequently, it can be used to define the
differential operator $\mathcal{Y}_{\ell}^{m} (\mathbf{\nabla})$ by
replacing in (\ref{YlmHomPol}) the Cartesian components $x$, $y$, and
$z$ of $\mathbf{r}$ by the Cartesian components $\partial / \partial x$,
$\partial / \partial y$, and $\partial / \partial z$ of
$\mathbf{\nabla}$.

For the integral of the product of three spherical harmonics over the
surface of the unit sphere in ${\mathbb{R}}^3$, the so-called Gaunt
coefficient \cite{Gau1929}, we write
\begin{eqnarray}
\lefteqn{\langle \ell_3 m_3 \vert \ell_2 m_2 \vert \ell_1 m_1 \rangle}
\nonumber \\
& = & \int \, \bigl[ Y_{\ell_3}^{m_3} (\Omega) \bigr]^{*} \, 
Y_{\ell_2}^{m_2} (\Omega) \,
Y_{\ell_1}^{m_1} (\Omega) \, {\mathrm{d}} \Omega \, . \quad
\end{eqnarray}
The Gaunt coefficient can be expressed in terms of $3jm$ symbols (see
for example p.\ 168 of \cite{ConOda1980}):
\begin{eqnarray}
\lefteqn{\langle \ell_3 m_3 \vert \ell_2 m_2 \vert \ell_1 m_1 \rangle} 
\nonumber \\
& = & (-1)^{m_3} \left[ \frac {(2\ell_1+1)(2\ell_2+1)(2\ell_3+1)}
{4\pi} \right]^{1/2} 
\nonumber \\
& \times & \! \!
\left( 
\begin{array}{ccc}
\ell_1 & \ell_2 & \ell_3 \\
0 & 0 & 0       
\end{array}  \right)
\left( 
\begin{array}{ccc}
\ell_1 & \ell_2 & \ell_3 \\
m_1 & m_2 & - m_3       
\end{array}  \right) \! . \qquad
\label{Gaunt_3jm}
\end{eqnarray}
It follows from the orthonormality of the spherical harmonics that the
Gaunt coefficients linearize the product of two spherical harmonics:
\begin{eqnarray}
\lefteqn{Y_{\ell_1}^{m_1} (\Omega) \, Y_{\ell_2}^{m_2} (\Omega)}
\nonumber \\
& = & \sum_{\ell=\ell_{\mathrm{min}}}^{\ell=\ell_{\mathrm{max}}} 
\! {}^{(2)} \,
\langle \ell m_1+m_2 \vert \ell_1 m_1 \vert \ell_2 m_2 \rangle 
\nonumber \\ 
& & \phantom{\sum_{\ell=\ell_{\mathrm{min}}}^{\ell=\ell_{\mathrm{max}}} 
\! {}^{(2)}} \, \times \, Y_{\ell}^{m_1+m_2} (\Omega) \, .
\label{Ylm_lin}
\end{eqnarray}
The symbol $\sum \! {}^{(2)}$ indicates that the summation proceeds in
steps of two. The summation limits in (\ref{Ylm_lin}), which follow from
the selection rules satisfied by the $3jm$ symbols, are given by
\cite{WeSt1982}
\begin{eqnarray}
\ell_{\mathrm{max}} & = & \ell_1 + \ell_2
\nonumber \\
\ell_{\mathrm{min}} & = & 
\begin{cases}
\lambda_{\mathrm{min}} \, , \\
\text{if $\ell_{\mathrm{max}} + \lambda_{\mathrm{min}}$ is even,} \\
{} \\
\lambda_{\mathrm{min}} + 1 \, , \\
\text{if $\ell_{\mathrm{max}} + \lambda_{\mathrm{min}}$ is odd,}
\end{cases}
\label{SumLims}
\end{eqnarray}
where
\begin{equation}
\lambda_{\mathrm{min}} \; = \; 
\max (\vert \ell_1 - \ell_2 \vert, m_1 + m_2) \, .
\end{equation}
It may be of interest to note that several articles on Gaunt
coefficients have appeared recently
\cite{HomSte1996,Xu1996,Xu1997,Xu1998,Seb1998,MavAla1999}.

In the sequel, the following abbreviations will be used:
\begin{eqnarray}
\Delta \ell & = & (\ell_1 + \ell_2 - \ell)/2 \, ,
\label{Def_Del_l} 
\\
\Delta \ell_1 & = & (\ell - \ell_1 + \ell_2)/2 \, ,
\label{Def_Del_l_1} 
\\
\Delta \ell_2 & = & (\ell + \ell_1 - \ell_2)/2 \, ,
\label{Def_Del_l_2} 
\\
\sigma (\ell) & = & (\ell_1 + \ell_2 + \ell)/2 \, .
\label{Def_sigma_l}
\end{eqnarray}
If the three orbital angular momentum quantum numbers $\ell_1$,
$\ell_2$, and $\ell$ satisfy the summation limits (\ref{SumLims}), then
these quantities are always positive integers or zero.

%
\setcounter{equation}{0}

\section{The Translation Operator in Spherical Form}
\label{Sec:TransOp}

In this Section, the expansion of the translation operator
$\mathrm{e}^{\mathbf{r}' \cdot \mathbf{\nabla}}$ in terms of irreducible
spherical tensors will be reviewed. Moreover, the for our purposes most
important properties of the differential operator
$\mathcal{Y}_{\ell}^{m} (\mathbf{\nabla})$, which is the key quantity in
the expansion mentioned above, will be discussed shortly. Additional
details be found in Sections 3 and 4 of \cite{Wen2000}.

If we want to derive an addition theorem for some function $f$ by
applying the translation operator according to (\ref{ExpDifOp}), $f$ has
to be \emph{analytic} at the expansion point $\mathbf{r}$ as well as at
the shifted argument $\mathbf{r} + \mathbf{r}'$. Consequently,
(\ref{ExpDifOp}) is in general \emph{not} symmetric with respect to an
interchange of $\mathbf{r}$ and $\mathbf{r}'$. Thus, if the
three-dimensional Taylor expansion (\ref{ExpDifOp}) converges, it is in
general \emph{not} possible to expand $f$ around $\mathbf{r}'$ and use
$\mathbf{r}$ as the shift vector according to
\begin{eqnarray}
f (\mathbf{r} + \mathbf{r}') & = & \sum_{n=0}^{\infty} \,
\frac{(\mathbf{r} \cdot \mathbf{\nabla}')^n}{n!} \, f (\mathbf{r}')
\nonumber \\
& = &
\mathrm{e}^{\mathbf{r} \cdot \mathbf{\nabla}'} \, f (\mathbf{r}') \, .
\label{ExpDifOpRev}
\end{eqnarray}   
The reason is that (\ref{ExpDifOp}) and (\ref{ExpDifOpRev}) can only
hold simultaneously for essentially arbitrary vectors $\mathbf{r},
\mathbf{r}' \in \mathbb{R}^3$ if $f$ is \emph{analytic} at $\mathbf{r}$, 
$\mathbf{r}'$, and at $\mathbf{r} + \mathbf{r}'$. However, most
functions, that are of interest in the context of atomic and molecular
quantum mechanics, are either singular at the origin or are not analytic
at the origin. Obvious examples are the Coulomb potential, which is
singular at the origin, or the $1 s$ hydrogen eigenfunction, which
possesses a cusp at the origin \cite{Kat1957}. In fact, all the commonly
used exponentially decaying functions as for example Slater-type
functions or also $B$ functions are not analytic at the origin.

The reason for the non-analyticity is that the three-dimensional
distance $r = \bigl[ x^2 + y^2 + z^2 \bigr]^{1/2}$ is not analytic with
respect to $x$, $y$, and $z$ at the origin $\mathbf{r} = \mathbf{0}$. In
contrast, $r^2 = x^2 + y^2 + z^2$ and the regular solid harmonic
$\mathcal{Y}_{\ell}^{m} (\mathbf{r})$ are analytic since they are
polynomial in $x$, $y$, and $z$. Consequently, a $1 s$ Gaussian function
$\exp (- \alpha r^2)$ is analytic at $\mathbf{r} = \mathbf{0}$, but a $1
s$ Slater-type function $\exp (- \alpha r)$ is not.

Thus, for the derivation of addition theorems for $B$ functions, which
are not analytic at the origin, we have to use the translation operator
in the following form,
\begin{eqnarray}
f (\mathbf{r}_{<} + \mathbf{r}_{>}) & = & 
\sum_{n=0}^{\infty} \,
\frac{(\mathbf{r}_{<} \cdot \mathbf{\nabla}_{>})^n}{n!} \, 
f (\mathbf{r}_{>}) \nonumber \\
& = & 
\mathrm{e}^{\mathbf{r}_{<} \cdot \mathbf{\nabla}_{>}} \, 
f (\mathbf{r}_{>}) \, , \quad
\label{TransOp<>}
\end{eqnarray}
where $0 \le \vert \mathbf{r}_{<} \vert < \vert \mathbf{r}_{>}
\vert$. In this way, the convergence of the Taylor expansion is
guaranteed provided that $f$ is analytical everywhere except at the
origin. Thus, the non-analyticity at the origin explains why $B$
functions as well as all the other commonly occurring exponentially
decaying functions have addition theorems with a two-range form of the
type of the Laplace expansion (\ref{LapExp}) of the Coulomb potential.

The crucial step, which ultimately makes the Taylor expansion method
practically useful, is the expansion of the translation operator
$\mathrm{e}^{\mathbf{r}_{<} \cdot \mathbf{\nabla}_{>}}$ in terms of
differential operators which are irreducible spherical tensors. The
starting point is an expansion of $\mathrm{e}^{\mathbf{a} \cdot
\mathbf{b}}$ with $\mathbf{a}, \mathbf{b} \in \mathbb{R}^3$ in terms 
of modified Bessel functions and Legendre polynomials (p.\ 108 of
\cite{MaObSo1966}):
\begin{eqnarray}
\lefteqn{\mathrm{e}^{\mathbf{a} \cdot \mathbf{b}} \; = \; 
\mathrm{e}^{a b \cos \theta} \; = \;
\left( \frac{\pi}{2 a b} \right)^{1/2}} \nonumber \\
& & \times \,  \sum_{\ell=0}^{\infty} \,
(2\ell+1) \, I_{\ell+1/2} (ab) \, P_{\ell} (\cos \theta) \, . \quad
\label{I_Pl_exp}
\end{eqnarray}
Next, the series expansion for the modified Bessel function
$I_{\ell+1/2}$ (p.\ 66 of \cite{MaObSo1966}) is inserted into
(\ref{I_Pl_exp}), and the Legendre polynomials are replaced by spherical
harmonics. This yields the following expansion of
$\mathrm{e}^{\mathbf{a} \cdot \mathbf{b}}$ in terms of regular solid
harmonics, which are irreducible spherical tensors of rank $\ell$, and
even powers of the vectors $\mathbf{a}$ and $\mathbf{b}$, which are
irreducible spherical tensors of rank zero:
\begin{eqnarray}
\mathrm{e}^{\mathbf{a} \cdot \mathbf{b}} & = &
2 \pi \, \sum_{\ell=0}^{\infty} \, \sum_{m=-\ell}^{\ell} \,
\left[ \mathcal{Y}_{\ell}^{m} (\mathbf{a}) \right]^{*} \, 
\mathcal{Y}_{\ell}^{m} (\mathbf{b})  \nonumber \\
& & \, \times \, \sum_{k=0}^{\infty} \, 
\frac{\mathbf{a}^{2k} \, \mathbf{b}^{2k}}
{2^{\ell+2k} k! (1/2)_{\ell+k+1}} \, .
\label{ST_Exp_e_ab} 
\end{eqnarray}
This relationship is obtained from (\ref{I_Pl_exp}) by rearranging the
Cartesian components of the vectors $\mathbf{a}$ and
$\mathbf{b}$. Accordingly, it holds for essentially arbitrary vectors
$\mathbf{a}$ and $\mathbf{b}$ and we can choose $\mathbf{a} =
\mathbf{r}_{<}$ and $\mathbf{b} = \mathbf{\nabla}_{>}$, yielding
\begin{eqnarray}
\mathrm{e}^{\mathbf{r}_{<} \cdot \mathbf{\nabla}_{>}} & = &
2 \pi \, \sum_{\ell=0}^{\infty} \, \sum_{m=-\ell}^{\ell} \,
\left[ \mathcal{Y}_{\ell}^{m} (\mathbf{r}_{<}) \right]^{*} \, 
\mathcal{Y}_{\ell}^{m} (\mathbf{\nabla}_{>})  \nonumber \\
& & \, \times \, \sum_{k=0}^{\infty} \, 
\frac{\mathbf{r}_{<}^{2k} \, \mathbf{\nabla}_{>}^{2k}}
{2^{\ell+2k} k! (1/2)_{\ell+k+1}} \, .
\label{ST_TransOp} 
\end{eqnarray}

It seems that (\ref{ST_TransOp}) was first published by Santos (Eq.\
(A.6) of \cite{San1973}). Santos also emphasized that this expansion
should be useful for the derivation of addition theorems, but apparently
did not use it for that purpose.

The only non-standard quantity in (\ref{ST_TransOp}) is the differential
operator $\mathcal{Y}_{\ell}^{m} (\mathbf{\nabla})$. Therefore, we will
now discuss those properties of $\mathcal{Y}_{\ell}^{m}
(\mathbf{\nabla})$ that are needed for the derivation of addition
theorems for $B$ functions. Other properties as well as numerous
applications -- mainly in the context of multicenter integrals -- are
described in the literature
\cite{Hob1965,WenSte1985,CleSchr1993,WenSte1989,Wen2000,%
CCHiMoh2001,San1973,Hob1892,Bay1978,Row1978,%
Mar1979,Fie1980,Stu1981,WenSte1983a,WenSte1983b,Nov1983,Niu1983,%
Niu1984,Niu1985,GroSte1985,WenGroSte1986,Ras1986,Nov1988,Hom1990,%
Dun1990,Mat1992a,Mat1992b,FujMat1992,ForCar1992,ForSal93,KuaLin1997,%
AraMat1998,CChMoh1998,CChMoh1999,BotMetKraSchm1998,HuStaBirIgoRoe2000,%
Dun2001}.

The spherical tensor gradient operator $\mathcal{Y}_{\ell}^{m}
(\mathbf{\nabla})$ is an irreducible spherical tensor of rank $\ell$
(p.\ 312 of \cite{BieLou1981}). Consequently, its application to a
function $\phi (r)$, which only depends on the distance $r$ and
therefore is an irreducible spherical tensor of rank zero, yields an
irreducible spherical tensor of rank $\ell$ according to
\begin{equation}
\mathcal{Y}_{\ell}^{m} (\mathbf{\nabla}) \, \phi (r) \; = \; 
\left[ 
\left( \frac{1}{r} \frac{\mathrm{d}}{\mathrm{d} r} \right)^{\ell} 
\phi (r) \right] \, \mathcal{Y}_{\ell}^{m} (\mathbf{r}) \, .
\label{TG_scal}
\end{equation}
As discussed in Section IV of \cite{WenSte1983a}, this relationship can
be derived with the help of a theorem on differentiation which was
published by Hobson already in the late 19th century
\cite{Hob1892}. Further details can be found on pp.\ 124 - 129 of
Hobson's book \cite{Hob1965} which was first published in 1931.

If the spherical tensor gradient operator is applied to a spherical
tensor of nonzero rank, i.e., to a function that can be written as
\begin{equation}
F_{\ell_2}^{m_2} (\mathbf{r}) \; = \; 
f_{\ell_2} (r) \, Y_{\ell_2}^{m_2} (\mathbf{r}/r) \, ,
\end{equation} 
the structure of the resulting expression can be understood in terms of 
angular momentum coupling (Eq.\ (3.9) of \cite{WenSte1983b}):
\begin{eqnarray}
\lefteqn{\mathcal{Y}_{\ell_1}^{m_1} (\mathbf{\nabla}) \, 
F_{\ell_2}^{m_2} (\mathbf{r})} \nonumber \\
& = & \sum_{\ell=\ell_{\mathrm{min}}}^{\ell=\ell_{\mathrm{max}}} 
\! {}^{(2)} \,
\langle \ell m_1+m_2 \vert \ell_1 m_1 \vert \ell_2 m_2 \rangle
\nonumber \\ 
& & \quad \times \, \gamma_{\ell_1 \ell_2}^{\ell} (r) \,
Y_{\ell}^{m_1 + m_2} (\mathbf{r}/r) \, .
\label{YlmFlm}
\end{eqnarray}
For the radial functions $\gamma_{\ell_1 \ell_2}^{\ell}$ in
(\ref{YlmFlm}), various representations could be derived, for example
(Eqs.\ (3.29) and (4.24) of \cite{WenSte1983b})
\begin{eqnarray}
\lefteqn{\gamma_{\ell_1 \ell_2}^{\ell} (r)} \nonumber \\
& = & \sum_{q=0}^{\Delta \ell} \, 
\frac{(-\Delta \ell)_q (-\sigma (\ell)-1/2)_q} {q!} \,
2^q \, r^{\ell_1 + \ell_2 - 2 q}
\nonumber \\
& & \times \, 
\left( \frac{1}{r} \frac{\mathrm{d}}{\mathrm{d} r} \right)^{\ell_1 - q}
\, \frac{f_{\ell_2} (r)}{r^{\ell_2}}
\label{gamma_1}
\\
& = & \sum_{s=0}^{\Delta \ell_2} \, 
\frac{(-\Delta \ell_2)_s (\Delta \ell_1 + 1/2)_s}{s!} \,
2^s \, r^{\ell_1 - \ell_2 - 2 s - 1}
\nonumber \\
& & \times
\left( \frac{1}{r} \frac{\mathrm{d}}{\mathrm{d} r} \right)^{\ell_1 - s}
\, r^{\ell_2 + 1} \, f_{\ell_2} (r) \, .
\label{gamma_2}
\end{eqnarray}
The abbreviations $\Delta \ell$, $\Delta \ell_1$, $\Delta \ell_2$, and
$\sigma (\ell)$ are defined in (\ref{Def_Del_l}) - (\ref{Def_sigma_l}).

These two as well as analogous other expressions for $\gamma_{\ell_1
\ell_2}^{\ell} (r)$ can be used for the derivation of addition theorems
of $B$ functions. However, in the case of $B$ functions more convenient
expressions are available. Consequently, the general expressions
(\ref{gamma_1}) and (\ref{gamma_2}) will actually not be used at all in
this article.

As mentioned before, $\mathcal{Y}_{\ell}^{m} (\mathbf{\nabla})$ is
obtained from the regular solid harmonic $\mathcal{Y}_{\ell}^{m}
(\mathbf{r})$ by replacing the Cartesian components $x$, $y$, and $z$ of
$\mathbf{r}$ by the Cartesian components $\partial/\partial x$,
$\partial/\partial y$, and $\partial/\partial z$ of
$\mathbf{\nabla}$. Consequently, $\mathcal{Y}_{\ell}^{m}
(\mathbf{\nabla})$ and $\mathcal{Y}_{\ell}^{m} (\mathbf{r})$ must obey
the same coupling law. Hence, (\ref{Ylm_lin}) implies (Eq.\ (6.24) of
\cite{WenSte1983b}):
\begin{eqnarray}
\lefteqn{\mathcal{Y}_{\ell_1}^{m_1} (\mathbf{\nabla}) \, 
\mathcal{Y}_{\ell_2}^{m_2} (\mathbf{\nabla})} \nonumber \\
& = & \sum_{\ell=\ell_{\mathrm{min}}}^{\ell=\ell_{\mathrm{max}}} 
\! {}^{(2)} \,
\langle \ell m_1+m_2 \vert \ell_1 m_1 \vert \ell_2 m_2 \rangle 
\nonumber \\ 
& & \, \times \, \mathbf{\nabla}^{\ell_1 + \ell_2 - \ell} \,
\mathcal{Y}_{\ell}^{m_1+m_2} (\mathbf{\nabla}) \, .
\label{STGO_cpl}
\end{eqnarray}
It follows from the summation limits (\ref{SumLims}) that the power
$\ell_1 + \ell_2 - \ell = 2 \Delta \ell$ of $\mathbf{\nabla}$ is either
zero or an even positive integer.

Let us now assume that a spherical tensor $F_{\ell_2}^{m_2}
(\mathbf{r})$ and a radially symmetric function $\Phi_{\ell_2} (r)$
exist which satisfy
\begin{equation}
F_{\ell_2}^{m_2} (\mathbf{r}) \; = \; 
\mathcal{Y}_{\ell_2}^{m_2} (\mathbf{\nabla}) \, \Phi_{\ell_2} (r) \, . 
\label{YlmNablaPhi}
\end{equation}
If we apply $\mathcal{Y}_{\ell_1}^{m_1} (\mathbf{\nabla})$ to
$F_{\ell_2}^{m_2} (\mathbf{r})$, then the two differential operators can
be coupled according to (\ref{STGO_cpl}). With the help of
(\ref{TG_scal}), we then obtain (Eq.\ (3.9) of \cite{WenSte1985}):
\begin{eqnarray}
\lefteqn{\mathcal{Y}_{\ell_1}^{m_1} (\mathbf{\nabla}) \,
F_{\ell_2}^{m_2} (\mathbf{r})} \nonumber \\ [1.5\jot]
& = & \mathcal{Y}_{\ell_1}^{m_1} (\mathbf{\nabla}) \, 
\mathcal{Y}_{\ell_2}^{m_2} (\mathbf{\nabla}) \, \Phi_{\ell_2} (r)
\nonumber \\
& = & \sum_{\ell=\ell_{\mathrm{min}}}^{\ell=\ell_{\mathrm{max}}} 
\! {}^{(2)} \,
\langle \ell m_1+m_2 \vert \ell_1 m_1 \vert \ell_2 m_2 \rangle \,
\nabla^{\ell_1 + \ell_2 - \ell}
\nonumber \\
& & \, \times \, \left[ 
\left( \frac{1}{r} \frac{\mathrm{d}}{\mathrm{d} r} \right)^{\ell} 
\Phi_{\ell_2} (r) \right] \,
\mathcal{Y}_{\ell}^{m_1+m_2} (\mathbf{r}) \, .
\label{YlmCplPhi}
\end{eqnarray}
This relationship is particularly well suited for $B$ functions. In this
case, (\ref{YlmCplPhi}) is more convenient than other, more general
expressions for the product $\mathcal{Y}_{\ell_1}^{m_1}
(\mathbf{\nabla}) F_{\ell_2}^{m_2} (\mathbf{r})$ which can for instance
be found in articles by Santos
\cite{San1973}, Bayman \cite{Bay1978}, Stuart \cite{Stu1981}, Niukkanen 
\cite{Niu1983}, Weniger and Steinborn \cite{WenSte1983b}, and Rashid 
\cite{Ras1986}.

%
\setcounter{equation}{0}

\section{$B$ Functions}
\label{Sec:BFun}

In this Section, those mathematical properties of $B$ functions, that
are relevant for the derivation of an addition theorem via
three-dimensional Taylor expansion, will be reviewed. More complete
treatments can be found in
\cite{WenSte1983a,FilSte1978b,FilSte1978a,Hom1990,Wen1982,%
WenSte1983c,WenGratSte1986,GroWenSte1986}.

If $K_{\nu} (z)$ is a modified Bessel function of the second kind (p.\
66 of \cite{MaObSo1966}), the reduced Bessel function is defined by
(Eqs.\ (3.1) and (3.2) of \cite{SteFil1975c})
\begin{equation}
\hat{k}_{\nu} (z) \; = \; (2/\pi)^{1/2} \, z^{\nu} \, K_{\nu} (z) \, .
\label{Def:RBF}
\end{equation}

If the order $\nu$ of a reduced Bessel function is half-integral, $\nu =
n + 1/2$ with $n \in \mathbb{N}_0$, the reduced Bessel function can be
written as an exponential multiplied a terminating confluent
hypergeometric series ${}_1 F_1$ (Eq.\ (3.7) of \cite{WenSte1983c}):
\begin{eqnarray}
\lefteqn{\hat{k}_{n+1/2} (z)} \nonumber \\
& = & 2^n \, (1/2)_n \, \mathrm{e}^{-z} \, {}_1 F_1 (-n; -2n; 2z) \, .
\quad
\label{RBF_HalfInt}
\end{eqnarray}
The polynomial part in (\ref{RBF_HalfInt}) was also treated
independently in the mathematical literature \cite{Gros1978}, where the
notation
\begin{equation}
\Theta_n (z) \; = \; \mathrm{e}^z \, \hat{k}_{n+1/2} (z) \, ,
\qquad n \in \mathbb{N}_0 \, ,
\end{equation}
is used. Together with some other, closely related polynomials, the
$\Theta_n (z)$ are called Bessel polynomials. They are applied in such
diverse fields as number theory, statistics, and the analysis of complex
electrical networks \cite{Gros1978}.

The so-called $B$ function was introduced by Filter and Steinborn as an
anisotropic generalization of the reduced Bessel function (Eq.\ (2.14)
of \cite{FilSte1978b}),
\begin{eqnarray}
\lefteqn{B_{n,\ell}^{m} (\alpha, \mathbf{r})} \nonumber \\
& = & [2^{n+\ell} (n+l)!]^{-1} \, \hat{k}_{n-1/2} (\alpha r) \, 
\mathcal{Y}_{\ell}^{m} (\alpha \mathbf{r}) \, , \qquad
\label{Def:B_Fun}
\end{eqnarray}
where $\alpha > 0$ and $n \in \mathbb{Z}$. Because of the factorial
$(n+\ell)!$ in the denominator, $B$ functions are defined in the sense
of classical analysis only if $n+\ell \ge 0$ holds. However, the
definition of a $B$ function remains meaningful even for $n+\ell <
0$. If $\mathbf{r} \neq \mathbf{0}$, $B_{-n-\ell,\ell}^{m} (\mathbf{r})$
with $n \in \mathbb{N}$ is zero, but for $\mathbf{r} = \mathbf{0}$, its
value is $\infty/\infty$ and therefore undefined. In fact, such a $B$
function can be interpreted as a derivative of the three-dimensional
Dirac delta function (Eq.\ (6.20) of \cite{WenSte1983b}):
\begin{eqnarray}
\lefteqn{B_{-n-\ell,\ell}^{m} (\mathbf{r}) \; = \; 
\frac{(2\ell-1)!! \, 4\pi}{\alpha^{\ell+3}}} \nonumber \\
&& \times \, 
\left[1 - \alpha^{-2} \mathbf{\nabla}^2 \right] \,
\delta_{\ell}^{m} (\mathbf{r}) \, , \quad n \in \mathbb{N} \, .
\end{eqnarray}
The spherical delta function $\delta_{\ell}^{m}$ is defined by
\begin{equation}
\delta_{\ell}^{m} (\mathbf{r}) \; = \; 
\frac{(-1)^{\ell}}{(2\ell-1)!!} \, 
\mathcal{Y}_{\ell}^{m} (\mathbf{\nabla}) \, \delta (\mathbf{r}) \, .
\end{equation}

If follows from (\ref{Def:RBF}) - (\ref{Def:B_Fun}) that $B$ functions
are relatively complicated mathematical objects. Nevertheless, all the
commonly occurring exponentially decaying functions can be expressed as
simple finite sums of $B$ functions. For example, if
\begin{equation}
\chi_{n, \ell}^{m} (\alpha, \mathbf{r}) \; = \; 
(\alpha r)^{n-1} \, \mathrm{e}^{- \alpha r} \, 
Y_{\ell}^{m} (\theta, \phi)
\label{Def:STF}
\end{equation}
is an (unnormalized) Slater-type function with $\alpha > 0$ and $n \ge
\ell + 1$, then (Eqs.\ (3.3) and (3.4) of \cite{FilSte1978b})
\begin{eqnarray}
\lefteqn{\chi_{n, \ell}^{m} (\alpha, \mathbf{r}) \; = \;
\sum_{p = p_\mathrm{min}}^{n-\ell} \, (-1)^{n-l-p}} \nonumber \\
&& \times 
\frac{(n-\ell)! 2^{\ell+p} (\ell+p)!} {(2p-n+\ell)! (2n-2\ell-2p)!!} \, 
B_{p,\ell}^{m} (\alpha, \mathbf{r}) \, , \qquad \;
\label{STF2B}
\end{eqnarray}
where $p_\mathrm{min} = \Ent {(n-\ell)/2}$. Here, $\Ent {x}$ stands for
the \emph{integral} part of $x$, i.e., for the largest integer $\nu$
satisfying $\nu \le x$.

As is well known, the bound-state eigenfunctions of a hydrogenlike ion
with nuclear charge $Z$ can be expressed as follows,
\begin{eqnarray}
\lefteqn{W_{n, \ell}^{m} (Z, \mathbf{r}) \; = \; 
\left( \frac{2Z}{n} \right)^{3/2} \, 
\left[ \frac{(n-\ell-1)!}{2n(n+\ell)!} \right]^{3/2} } \nonumber \\
& & \times \, \mathrm{e}^{-Zr/n} \, L_{n-\ell-1}^{(2\ell+1)} (2Zr/n) \,
\mathcal{Y}_{\ell}^{m} (2Z \mathbf{r}/n) \, , \qquad
\label{Def:HydEigFun}
\end{eqnarray}
where $L_{n-\ell-1}^{(2\ell+1)}$ is a generalized Laguerre
polynomial. Then (Eq.\ (3.16) of \cite{WenSte1989}),
\begin{eqnarray}
\lefteqn{W_{n, \ell}^{m} (Z, \mathbf{r}) \; = \; 
\left( \frac{2Z}{n} \right)^{3/2}} \nonumber \\ 
&& \times \, \frac{2^{\ell+1}}{(2\ell+1)!!} \, 
\left[ \frac{n(n+\ell)!}{2(n-\ell-1)!} \right]^{1/2} \nonumber \\
&& \quad \times \, \sum_{t=0}^{n-\ell-1} \, 
\frac{(-n+\ell+1)_t (n+\ell+1)_t} {t!(\ell+3/2)_t} \nonumber \\
&& \qquad \times \, B_{t+1,\ell}^{m} (Z/n, \mathbf{r}) \, .
\label{HydEigFun2B}
\end{eqnarray}
The bound-state hydrogenic eigenfunctions are orthonormal, but not
complete in the Hilbert space $L^2 (\mathbb{R}^3)$. This incompleteness
is sometimes overlooked (see for example the discussion in
\cite{WenSte1984}).

The following set of functions, which was introduced by Hylleraas
\cite{Hyl1929} and by Shull and L\"owdin \cite{ShuLow1955,LowShu1956},
is complete and orthonormal in the Hilbert space $L^2 (\mathbb{R}^3)$:
\begin{eqnarray}
\lefteqn{\Lambda_{n, \ell}^{m} (\alpha, \mathrm{R}) \; = \;
(2 \alpha)^{3/2} \frac{(n-\ell-1)!}{(n+\ell+1)!}} \nonumber \\
&& \times \, \mathrm{e}^{-\alpha r} \, 
L_{n-\ell-1}^{(2\ell+2)} (2 \alpha r) \,
\mathcal{Y}_{\ell}^{m} (2 \alpha \mathrm{r}) \, .
\label{Def:Lambda}
\end{eqnarray}
Then (Eq.\ (3.18) of \cite{FilSte1980}),
\begin{eqnarray}
\lefteqn{\Lambda_{n, \ell}^{m} (\alpha, \mathbf{r}) \; = \; 
(2 \alpha)^{3/2}} \nonumber \\ 
&& \times \, 2^{\ell} \, \frac{(2n+1)}{(2\ell+3)!!} \, 
\left[ \frac{(n+\ell+1)!}{(n-\ell-1)!} \right]^{1/2} \nonumber \\
&& \quad \times \, \sum_{t=0}^{n-\ell-1} \, 
\frac{(-n+\ell+1)_t (n+\ell+2)_t} {t!(\ell+5/2)_t} \nonumber \\
&& \qquad \times \, B_{t+1,\ell}^{m} (\alpha, \mathbf{r}) \, .
\label{Lambda2B}
\end{eqnarray}

Closely related to the bound-state hydrogenic eigenfunctions is the
following set of functions which were already used in 1928 by Hylleraas
\cite{Hyl1928} and which are commonly called Coulomb Sturmians or simply
Sturmians
\cite{Rot1970}:
\begin{eqnarray}
\lefteqn{\Psi_{n, \ell}^{m} (\alpha, \mathbf{r}) \; = \; 
(2 \alpha)^{3/2} \, 
\left[ \frac{(n-\ell-1)!}{2n(n+\ell)!} \right]^{3/2} } \nonumber \\
& & \times \, \mathrm{e}^{-\alpha r} \, L_{n-\ell-1}^{(2\ell+1)} (2
\alpha r) \, \mathcal{Y}_{\ell}^{m} (2 \alpha \mathbf{r}) \, . \quad
\label{Def:Psi}
\end{eqnarray} 
Comparison of (\ref{Def:HydEigFun}) and (\ref{Def:Psi}) yields
\begin{equation}
\Psi_{n, \ell}^{m} (Z/n, \mathbf{r}) \; = \;
W_{n, \ell}^{m} (Z, \mathbf{r}) \, .
\end{equation}
The fact, that bound-state hydrogenic eigenfunctions and Sturmians have
the same normalization constant, is actually by no means obvious. The
$W_{n, \ell}^{m} (Z, \mathbf{r})$ are normalized with respect to norm of
the Hilbert space $L^2 (\mathbb{R}^3)$, whereas the $\Psi_{n, \ell}^{m}
(\alpha, \mathbf{r})$ are normalized with respect to the norm of the
Sobolev space $W_{2}^{(1)} (\mathbb{R}^3)$ \cite{Sob1963}. Further
details can be found in
\cite{Wen1985}.

In (\ref{HydEigFun2B}), we only have to replace $Z/n$ by $\alpha$
to obtain the following representation of Sturmians in terms of $B$
functions (Eq.\ (4.19) of \cite{Wen1985}): 
\begin{eqnarray}
\lefteqn{\Psi_{n, \ell}^{m} (\alpha, \mathbf{r}) \; = \; 
(2 \alpha)^{3/2}} \nonumber \\ 
&& \times \, \frac{(2^{\ell+1}}{(2\ell+1)!!} \, 
\left[ \frac{n(n+\ell)!}{2(n-\ell-1)!} \right]^{1/2} \nonumber \\
&& \quad \times \, \sum_{t=0}^{n-\ell-1} \, 
\frac{(-n+\ell+1)_t (n+\ell+1)_t} {t!(\ell+3/2)_t} \nonumber \\
&& \qquad \times \, B_{t+1,\ell}^{m} (\alpha, \mathbf{r}) \, .
\label{Psi2B}
\end{eqnarray}
The linear combinations (\ref{STF2B}), (\ref{HydEigFun2B}),
(\ref{Lambda2B}), and (\ref{Psi2B}) imply that mathematical results for
$B$ functions can be translated immediately to analogous results for
Slater-type functions, bound-state hydrogenic eigenfunctions, Lambda
functions, and Sturmians.

In view of the fact that $B$ functions have a relatively complicated
mathematical structure, it is by no means trivial that the linear
combinations (\ref{STF2B}), (\ref{HydEigFun2B}), (\ref{Lambda2B}), and
(\ref{Psi2B}) could be derived at all. However, these relationships as
well as other advantageous properties of $B$ functions can be explained
via their Fourier transform, which is is of exceptional simplicity among
exponentially declining functions (Eq.\ (7.1-6) of \cite{Wen1982} or
Eq.\ (3.7) of \cite{WenSte1983a}):
\begin{eqnarray}
\lefteqn{\bar{B}_{n,\ell}^{m} (\mathbf{p}) \; = \; (2\pi)^{-3/2} \, \int \,
\mathrm{e}^{- i \mathbf{p} \cdot \mathbf{r}} \,
B_{n,\ell}^{m} (\mathbf{r}) \, \mathrm{d}^3 \mathbf{r}} \nonumber \\
& = & (2/\pi)^{1/2} \,
\frac{\alpha^{2n+\ell-1}}{[\alpha^2 + p^2]^{n+\ell+1}} \,
\mathcal{Y}_{\ell}^{m} (- i \mathbf{p}) \, . \qquad
\label{FT_B_Fun}
\end{eqnarray}        
The linear combinations (\ref{STF2B}), (\ref{HydEigFun2B}),
(\ref{Lambda2B}), and (\ref{Psi2B}) imply that the Fourier transforms of
Slater-type functions, of the bound-state hydrogenic eigenfunctions, of
Lambda functions, and of Sturmians can be expressed as simple linear
combinations of Fourier transforms of $B$ functions.

If we want to derive an addition theorem for some function via the
translation operator in its spherical form (\ref{ST_TransOp}), then the
application of higher powers of the Laplacian and of the spherical
tensor gradient operator to this function must lead to expressions of
manageable complexity. On the basis of the Fourier transform
(\ref{FT_B_Fun}), it can be seen that these expressions are indeed very
simple in the case of $B$ functions.

The differential operator $1 - \mathbf{\nabla}^2/\alpha^2$, which is
typical of the modified Helmholtz equation, acts on $B$ functions as a
ladder operator for the order $n$ (Eq.\ (5.6) of \cite{WenSte1983b}):
\begin{equation}
[1 - \mathbf{\nabla}^2/\alpha^2] \, B_{n,\ell}^{m} (\mathbf{r}) \; = \;
B_{n-1,\ell}^{m} (\mathbf{r}) \, .
\label{Bnlm_Ladder}
\end{equation}
This follows at once from the Fourier transform (\ref{FT_B_Fun}) in
combination with the fact that in momentum space the differential
operator $[1 - \mathbf{\nabla}^2/\alpha^2]$ is transformed into the
multiplicative operator $[\alpha^2 + p^2]/\alpha^2$.

With the help of this relationship, the application of the higher powers
of the Laplacian $\mathbf{\nabla}^{2}$ can be expressed quite easily in
closed form. Combination of (\ref{Bnlm_Ladder}) with the binomial
theorem yields (Eq.\ (5.7) of \cite{WenSte1983b}):
\begin{eqnarray}
\lefteqn{\alpha^{-2\nu} \, \mathbf{\nabla}^{2\nu} \,
B_{n,\ell}^{m} (\mathbf{r})} \nonumber \\
& = & \sum_{t=0}^{\nu} \, (-1)^t \, {\binom{\nu} {t}} \,
B_{n-t,\ell}^{m} (\mathbf{r}) \, .
\label{Delta2nB_Fun}
\end{eqnarray}        

The remarkably simple Fourier transform (\ref{FT_B_Fun}) also explains
why the application of the spherical tensor gradient operator to a $B$
function produces particularly compact expressions. For example, if we
combine
\begin{equation}
\mathcal{Y}_{\ell}^{m} (\mathbf{\nabla}) \,
\mathrm{e}^{\pm i \mathbf{p} \cdot \mathbf{r}} \; = \;
\mathcal{Y}_{\ell}^{m} (\pm i \mathbf{\mathbf{p}})
\mathrm{e}^{\pm i \mathbf{p} \cdot \mathbf{r}}
\end{equation}
with the Fourier transform (\ref{FT_B_Fun}), we find that a
\emph{nonscalar} $B$ function can be obtained by applying the
corresponding spherical tensor gradient operator to a \emph{scalar} $B$
function (Eq.\ (4.12) of \cite{WenSte1983a}):
\begin{equation}
B_{n,\ell}^{m} (\mathbf{r}) \; = \;
\frac{(4\pi)^{1/2}}{(-\alpha)^{\ell}}  \,  \,
\mathcal{Y}_{\ell}^{m} (\mathbf{\nabla}) \,
B_{n+\ell,0}^{0} (\mathbf{r}) \, .
\label{STGO_Bn00}
\end{equation}
This relationship can also be obtained in a straightforward way with the
help of (\ref{TG_scal}). The differential operator $\bigl(r^{-1}
\mathrm{d}/ \mathrm{d} r \bigr)^{\ell}$ in (\ref{TG_scal}) is a simple
operator for modified Bessel functions, but not for the other
exponentially decaying functions considered in this article. It acts as
a shift operator for reduced Bessel functions according to
\begin{equation}
\left( \frac{1}{z} \frac{\mathrm{d}}{\mathrm{d} z} \right)^m \,
\hat{k}_{\nu} (z) \; = \; (-1)^m \, \hat{k}_{\nu-m} (z) \, .
\end{equation}

Similarly, the application of the spherical tensor gradient operator to
a \emph{nonscalar} $B$ function with $\ell > 0$ yields a simple linear
combination of $B$ functions (Eq.\ (6.25) of \cite{WenSte1983b}):
\begin{eqnarray}
\lefteqn{\mathcal{Y}_{\ell_1}^{m_1} (\mathbf{\nabla}) \,
B_{n_2,\ell_2}^{m_2} (\mathbf{r})} \nonumber \\
& = & (-\alpha)^{\ell_1} \,
\sum_{\ell=\ell_{\mathrm{min}}}^{\ell=\ell_{\mathrm{max}}}
\! {}^{(2)} \,
\langle \ell m_1+m_2 \vert \ell_1 m_1 \vert \ell_2 m_2 \rangle
\nonumber \\
& & \, \times \, \sum_{t=0}^{\Delta \ell} \, (-1)^t \,
{\binom{\Delta \ell} {t}} \,
B_{n_2+\ell_2-\ell-t,\ell}^{m_1+m_2} (\mathbf{r}) \, .
\label{STGO_Bnlm}
\end{eqnarray}
This relationship follows at once from (\ref{YlmCplPhi}),
(\ref{Delta2nB_Fun}), and (\ref{STGO_Bn00}).

%
\setcounter{equation}{0}

\section{An Addition Theorem for $B$ Functions}
\label{Sec:AddTheorBFun}

We now want to derive an addition theorem for $B$ functions via
three-dimensional Taylor expansion. For that purpose, we apply the
translation operator in its spherical form (\ref{ST_TransOp}) to a $B$
function:
\begin{eqnarray}
\lefteqn{B_{n, \ell}^{m} (\alpha, \mathbf{r}_{<} + \mathbf{r}_{>})}
\nonumber \\
& = & \! 2 \pi \sum_{\ell_1=0}^{\infty} \, \sum_{m = - \ell_1}^{\ell_1}
\, \left[ \mathcal{Y}_{\ell_1}^{m_1} (\mathbf{r}_{<}) \right]^{*}
\mathcal{Y}_{\ell_1}^{m_1} (\mathbf{\nabla}_{>}) \nonumber \\
& & \! \times \, \sum_{q=0}^{\infty} \,
\frac{r_{<}^{2q} \, \mathbf{\nabla}_{>}^{2q}}
{2^{\ell_1+2q} q! (1/2)_{\ell_1+q+1}} \,
B_{n ,\ell}^{m} (\alpha, \mathbf{r}_{>}) \, . \qquad
\label{AddBFun_1}
\end{eqnarray}      
Now, we want to apply the powers of the Laplacian
$\mathbf{\nabla}_{>}^{2}$ to the $B$ functions in
(\ref{AddBFun_1}). This can be done with the help of
(\ref{Delta2nB_Fun}), yielding
\begin{eqnarray}
\lefteqn{B_{n, \ell}^{m} (\alpha, \mathbf{r}_{<} + \mathbf{r}_{>})}
\nonumber \\
& = & \! 2 \pi \sum_{\ell_1=0}^{\infty} \, \sum_{m=- \ell_1}^{\ell_1}
\, \left[ \mathcal{Y}_{\ell_1}^{m_1} (\mathbf{r}_{<}) \right]^{*}
\mathcal{Y}_{\ell_1}^{m_1} (\mathbf{\nabla}_{>}) \nonumber \\
& & \! \times \, \sum_{q=0}^{\infty} \,
\frac{(\alpha r_{<})^{2q}}
{2^{\ell_1+2q} q! (1/2)_{\ell_1+q+1}} \nonumber \\
& & \quad \times \, \sum_{t=0}^{q} \, (-1)^t \, {\binom {q}{t}} \,
B_{n - t,\ell}^{m} (\alpha, \mathbf{r}_{>}) \, . \qquad
\label{AddBFun_2}
\end{eqnarray}      
It follows from the definition of a $B$ function according to
(\ref{Def:B_Fun}) that $B_{n, \ell}^{m} (\mathbf{r}_{>})$ with
$\mathbf{r}_{>} \ne \mathbf{0}$ is zero for $n < - \ell$. Thus, the
summation limit of the innermost summation in (\ref{AddBFun_2}) is not
$q$ but $\min (q, n+\ell)$. Hence, if we introduce a new summation
variable $s = q - t$, we obtain for the two innermost summations in
(\ref{AddBFun_2}):
\begin{eqnarray}
\lefteqn{\sum_{q=0}^{\infty} \,
\frac{(\alpha r_{<})^{2q}}
{2^{\ell_1+2q} q! (1/2)_{\ell_1+q+1}}} \nonumber \\
& & \times \, \sum_{t=0}^{q} \, (-1)^t \, {\binom {q}{t}} \,
B_{n - t,\ell}^{m} (\alpha, \mathbf{r}_{>}) \nonumber \\
& = & \sum_{q=0}^{\infty} \,
\frac{(\alpha r_{<})^{2q}}
{2^{\ell_1+2q} q! (1/2)_{\ell_1+q+1}} \nonumber \\
& & \times \, 
\sum_{t=0}^{\min (q, n+\ell)} \, (-1)^t \, {\binom {q}{t}} \,
B_{n - t,\ell}^{m} (\alpha, \mathbf{r}_{>}) \qquad
\\
& = & 2^{-\ell_1} \, 
\sum_{t=0}^{n+\ell} \, \frac{(-1)^t}{t!} \,
B_{n - t,\ell}^{m} (\alpha, \mathbf{r}_{>}) \nonumber \\
& & \times \, \sum_{s=0}^{\infty} \, 
\frac{(\alpha r_{<}/2)^{2s+2t}}{s! (1/2)_{\ell_1+s+t+1}} 
\\
& = & 2^{-\ell_1} \, 
\sum_{t=0}^{n+\ell} \, 
\frac {(-1)^t (\alpha r_{<}/2)^{2t}} {t! (1/2)_{\ell_1+t+1}} \, 
B_{n - t,\ell}^{m} (\alpha, \mathbf{r}_{>}) \nonumber \\
& & \quad \times \, 
{}_0 F_1 \bigl( \ell_1+t+3/2; (\alpha r_{<})^2/4 \bigr) \, .
\label{InnSum}
\end{eqnarray}
The generalized hypergeometric series ${}_0 F_1$ can be replaced by a
modified Bessel function of the first kind according to (p.\ 66 of
\cite{MaObSo1966}) 
\begin{equation}
{}_0 F_1 (\nu+1; z^2/4) \; = \; 
\frac{\Gamma (\nu+1)}{(z/2)^{\nu}} \, I_{\nu} (z) \, .
\label{0F1_2_I}
\end{equation}
By combining (\ref{InnSum}) and (\ref{0F1_2_I}) we obtain
\begin{eqnarray}
\lefteqn{\sum_{q=0}^{\infty} \,
\frac{(\alpha r_{<})^{2q}}
{2^{\ell_1+2q} q! (1/2)_{\ell_1+q+1}}} \nonumber \\
& & \times \, \sum_{t=0}^{q} \, (-1)^t \, {\binom {q}{t}} \,
B_{n - t,\ell}^{m} (\alpha, \mathbf{r}_{>}) \nonumber \\
& = & \pi^{-1/2} \, 
\sum_{t=0}^{n+\ell} \, 
\frac {(-1)^t 2^{1/2-t}} {t! (\alpha r_{<})^{\ell_1-t+1/2}} \, 
I_{\ell_1+t+1/2} (\alpha r_{<}) \nonumber \\
& & \qquad \qquad \times \, 
B_{n - t,\ell}^{m} (\alpha, \mathbf{r}_{>}) \, .
\label{InnSumI}
\end{eqnarray}
Inserting (\ref{InnSumI}) into (\ref{AddBFun_2}) yields: 
\begin{eqnarray}
\lefteqn{B_{n, \ell}^{m} (\alpha, \mathbf{r}_{<} + \mathbf{r}_{>})}
\nonumber \\
& = & \! (2\pi)^{3/2} 
\sum_{\ell_1=0}^{\infty} \, \sum_{m=- \ell_1}^{\ell_1}
\, \left[ \mathcal{Y}_{\ell_1}^{m_1} (\mathbf{r}_{<}) \right]^{*}
\mathcal{Y}_{\ell_1}^{m_1} (\mathbf{\nabla}_{>}) \nonumber \\
& & \! \times \, \sum_{t=0}^{n+\ell} \,
\frac {(-1)^t} {2^t t!} \, (\alpha r_{<})^{t-\ell_1-1/2} \,
I_{\ell_1+t+1/2} (\alpha r_{<}) \nonumber \\
& & \qquad \times \, 
B_{n - t,\ell}^{m} (\alpha, \mathbf{r}_{>}) \, .
\label{AddBFun_3}
\end{eqnarray}      
Now, all that remains to be done is the application of the spherical
tensor gradient operator to the $B$ functions. This can be done with the
help of (\ref{STGO_Bnlm}), yielding
\begin{eqnarray}
\lefteqn{\mathcal{Y}_{\ell_1}^{m_1} (\mathbf{\nabla}_{>}) \,
B_{n-t,\ell}^{m} (\mathbf{r}_{>})} \nonumber \\
& = & (-\alpha)^{\ell_1} \,
\sum_{\ell_2=\ell_2^{\mathrm{min}}}^{\ell_2=\ell_2^{\mathrm{max}}}
\! {}^{(2)} \,
\langle \ell_2 m+m_1 \vert \ell_1 m_1 \vert \ell m \rangle
\nonumber \\
& & \, \times \, \sum_{s=0}^{\Delta \ell_2} \, (-1)^s \,
{\binom{\Delta \ell_2} {s}} \,
B_{n+\ell-\ell_2-s-t,\ell_2}^{m+m_1} (\mathbf{r}_{>}) \qquad \, .
\label{LastInnSum}
\end{eqnarray}
The summation limit $\Delta \ell_2$ in (\ref{LastInnSum}) is defined in
(\ref{Def_Del_l_2}). By inserting (\ref{LastInnSum}) into
(\ref{AddBFun_3}), we finally obtain an addition theorem for $B$
functions: 
\begin{eqnarray}
\lefteqn{B_{n, \ell}^{m} (\alpha, \mathbf{r}_{<} + \mathbf{r}_{>})}
\nonumber \\
& = & \! (2\pi)^{3/2} 
\sum_{\ell_1=0}^{\infty} \, \sum_{m=- \ell_1}^{\ell_1}
\, (-1)^{\ell_1} \,
\, \left[ \mathcal{Y}_{\ell_1}^{m_1} (\mathbf{r}_{<}) \right]^{*} 
\nonumber \\
& \times & \sum_{t=0}^{n+\ell} \,
\frac {(-1)^t} {2^t t!} \, (\alpha r_{<})^{t-\ell_1-1/2} \,
I_{\ell_1+t+1/2} (\alpha r_{<}) \nonumber \\
& \times & 
\sum_{\ell_2=\ell_2^{\mathrm{min}}}^{\ell_2=\ell_2^{\mathrm{max}}}
\! {}^{(2)} \,
\langle \ell_2 m+m_1 \vert \ell_1 m_1 \vert \ell m \rangle
\nonumber \\
& \times & \sum_{s=0}^{\Delta \ell_2} \, (-1)^s \,
{\binom{\Delta \ell_2} {s}} \,
B_{n+\ell-\ell_2-s-t,\ell_2}^{m+m_1} (\mathbf{r}_{>}) \, . \; \qquad
\label{AddBFun_4}
\end{eqnarray}
This expression is not yet completely satisfactory from a formal point
of view. We have to take into account that the $B$ functions in the
innermost sum are zero if $n+\ell-\ell_2-s-t < - \ell_2$. For that
purpose, we introduce a new summation variable $q = n+\ell-t$. This
yields:
\begin{eqnarray}
\lefteqn{B_{n, \ell}^{m} (\alpha, \mathbf{r}_{<} + \mathbf{r}_{>}) 
\; = \; \frac{(2\pi)^{3/2}}{(-2)^{n+\ell}}} \nonumber \\
& \times &  
\sum_{\ell_1=0}^{\infty} \, \sum_{m=- \ell_1}^{\ell_1}
\, (-1)^{\ell_1} \,
\, \left[ \mathcal{Y}_{\ell_1}^{m_1} (\mathbf{r}_{<}) \right]^{*} 
\nonumber \\
& \times & \sum_{q=0}^{n+\ell} \, \frac {(-2)^q} {(n+\ell-q)!} \, 
(\alpha r_{<})^{n+\ell-\ell_1-q-1/2}
\nonumber \\
& & \qquad \times \,
I_{n+\ell+\ell_1-q+1/2} (\alpha r_{<}) \nonumber \\
& \times & 
\sum_{\ell_2=\ell_2^{\mathrm{min}}}^{\ell_2=\ell_2^{\mathrm{max}}}
\! {}^{(2)} \,
\langle \ell_2 m+m_1 \vert \ell_1 m_1 \vert \ell m \rangle
\nonumber \\
& \times & \sum_{s=0}^{\min (q, \Delta \ell_2)} \, (-1)^s \,
{\binom{\Delta \ell_2} {s}} \nonumber \\
& & \qquad \times \,
B_{q-\ell_2-s,\ell_2}^{m+m_1} (\mathbf{r}_{>}) \, . \qquad
\label{AddBFun_5}
\end{eqnarray}      

This addition theorem is identical to the addition theorem (4.27) of
\cite{WenSte1989} which was derived with the help of techniques that 
closely resemble the zeta function method of Barnett and
Coulson \cite{BarCou1951}. Other addition theorems for $B$ functions and 
for reduced Bessel functions can be found in Sections IV and V of
\cite{WenSte1989}. 
 
%
\setcounter{equation}{0}

\section{Summary and Conclusions}
\label{Sec:SumCon}

Addition theorems are expansions of a given function $f (\mathbf{r} \pm
\mathbf{r}')$ with $\mathbf{r}, \mathbf{r}' \in \mathbb{R}^3$ in terms
of other functions that only depend on either $\mathbf{r}$ or
$\mathbf{r}'$. As discussed in Section \ref{Sec:Intro}, many different
techniques for the construction of addition theorems are known. Examples
are the zeta function method of Barnett and Coulson \cite{BarCou1951},
L\"{o}wdin's alpha function method \cite{Low1956}, or the Fourier
transform method suggested independently by Ruedenberg \cite{Rue1967}
and Silverstone\cite{Sil1967}.

Addition theorems can also be defined as three-dimensional Taylor
expansions according to (\ref{ExpDifOp}). In this approach, the
translation operator $\mathrm{e}^{\mathbf{r}' \cdot \mathbf{\nabla}}$
generates $f (\mathbf{r} + \mathbf{r}')$ by doing a three-dimensional
Taylor expansion of $f$ around $\mathbf{r}$. In the context of atomic
and molecular calculations, one is normally interested in addition
theorems of irreducible spherical tensors of the type of
(\ref{IrrSpheTen}). In such a case, the straightforward application of
the translation operator in its Cartesian form $\mathrm{e}^{x' \partial
/\partial x} \mathrm{e}^{y' \partial /\partial y} \mathrm{e}^{z'
\partial /\partial z}$ would lead to enormous technical
problems. Consequently, it was generally believed that three-dimensional
Taylor expansions constitute essentially a formal solution, but cannot
be used for the actual derivation of addition theorems of irreducible
spherical tensors. However, the Taylor expansion method is indeed a
practically useful tool if the translation operator is expanded in terms
of differential operators that are irreducible spherical tensors
according to (\ref{ST_TransOp}). This was demonstrated in
\cite{Wen2000} by constructing the Laplace expansion of the Coulomb
potential as well as the addition theorems of the regular and irregular
solid harmonics and of the Yukawa potential.

Since the addition theorems mentioned above are all comparatively
simple, their successful construction does not necessarily guarantee
that the Taylor expansion method is also practically useful if more
complicated addition theorems are to be constructed. Consequently, the
topic of the present article is the construction of addition theorems
for exponentially decaying functions which are notoriously complicated.

Slater-type functions have the simplest structure of all the commonly
occurring exponentially decaying functions. Consequently it looks like
an obvious idea to concentrate on addition theorems of Slater-type
functions, and to express the addition theorems for other exponentially
decaying functions in terms of addition theorems for Slater-type
functions.

However, the addition theorems for Slater-type functions are notoriously
difficult. Moreover, Slater-type functions have a simple analytical
structure only in the coordinate representation. In momentum space,
which is probably more important in the case of multicenter problems,
the simplest exponentially decaying functions are the so-called $B$
functions. Their extremely simple Fourier transform (\ref{FT_B_Fun}) as
well as some other advantageous mathematical properties give $B$
functions a unique position among exponentially decaying
functions. Moreover, as discussed in Section \ref{Sec:BFun}, all the
commonly occurring exponentially decaying functions can be expressed as
simple finite sums of $B$ functions. This implies that the addition
theorems of Slater-type functions, of bound-state hydrogenic
eigenfunctions, and of other functions based on Laguerre polynomials can
be expressed in terms of the addition theorems of $B$ functions.

So, this article focuses on the derivation of an addition theorem for
$B$ functions by doing a three-dimensional Taylor expansion in its
spherical form (\ref{ST_TransOp}). The relative ease, with which the
addition theorem (\ref{AddBFun_5}) can be constructed, and its
comparatively compact structure indicate once more that $B$ functions
have in the context of multicenter problems considerable advantages over
other exponentially decaying functions.

\typeout{==> References}

\end{multicols}

\end{document}